\documentclass[]{spie}  

\usepackage{amsmath,amsfonts,amssymb}
\usepackage{graphicx}
\usepackage[colorlinks=true, allcolors=blue]{hyperref}
\usepackage{threeparttable}
\usepackage{booktabs}
\usepackage{multirow}
\usepackage{gensymb}
\usepackage{makecell}
\usepackage{longtable}
\usepackage{bm}

\title{ISIT-GEN: An in silico imaging trial to assess the inter-scanner generalizability of CTLESS for myocardial perfusion SPECT on defect-detection task}

\author[a]{Zitong Yu}
\author[b]{Nu Ri Choi}
\author[c]{Zezhang Yang}
\author[d]{Nancy A. Obuchowski}
\author[b]{Barry A. Siegel}
\author[a,b,c]{Abhinav K. Jha}
\affil[a]{Department of Biomedical Engineering, Washington University, St. Louis, MO, USA}
\affil[b]{Mallinckrodt Institute of Radiology, Washington University, St. Louis, MO, USA}
\affil[c]{Department of Electrical and Systems Engineering, Washington University, St. Louis, MO, USA}
\affil[d]{Department of Quantitative Health Sciences, Cleveland Clinic, Cleveland, OH, USA}

\authorinfo{Further author information: (Send correspondence to A. K. Jha)\\A. K. Jha: E-mail: a.jha@wustl.edu; Z. Yu: E-mail: yu.zitong@wustl.edu;
N. R. Choi: E-mail: nuri.choi@wustl.edu; Z. Yang: E-mail: y.zezhang@wustl.edu; B. A. Siegel: E-mail: siegelb@wustl.edu}

\begin{document} 
This manuscript has been accepted to SPIE Medical Imaging, February 16-20, 2025. Please use the following reference when citing the manuscript.

Yu, Z., Choi, N., Yang, Z, Obuchowski, N. A., Siegel, B. A., and Jha, A. K., “ISIT-GEN: An in silico imaging trial to assess the inter-scanner generalizability of CTLESS for myocardial perfusion SPECT on defect-detection task”, Proc. SPIE Medical Imaging, 2025.

\newpage
\maketitle

\begin{abstract}
A recently proposed scatter-window and deep learning-based attenuation compensation (AC) method for myocardial perfusion imaging (MPI) by single-photon emission computed tomography (SPECT), namely CTLESS, demonstrated promising performance on the clinical task of myocardial perfusion defect detection with retrospective data acquired on SPECT scanners from a single vendor. For clinical translation of CTLESS, it is important to assess the generalizability of CTLESS across different SPECT scanners. For this purpose, we conducted a virtual imaging trial, titled \textit{\underline{i}n \underline{s}ilico} \underline{i}maging \underline{t}rial to assess \underline{gen}eralizability (ISIT-GEN). ISIT-GEN assessed the generalizability of CTLESS on the cardiac perfusion defect detection task across SPECT scanners from three different vendors. The performance of CTLESS was compared with a standard-of-care CT-based AC (CTAC) method and a no-attenuation compensation (NAC) method using an anthropomorphic model observer. We observed that CTLESS had receiver operating characteristic (ROC) curves and area under the ROC curves similar to those of CTAC. Further, CTLESS was observed to significantly outperform the NAC method across three scanners. These results are suggestive of the inter-scanner generalizability of CTLESS and motivate further clinical evaluations. The study also highlights the value of using \textit{in silico} imaging trials to assess the generalizability of deep learning-based AC methods feasibly and rigorously.
\end{abstract}

\keywords{\textit{In silico} imaging trial, transmission-less attenuation compensation, deep learning, generalizability, myocardial perfusion SPECT}

\section{INTRODUCTION}
\label{sec:intro}  
Coronary artery disease is the single largest cause of death worldwide, and also a major cause of hospital admissions. Myocardial perfusion imaging (MPI) by single-photon emission computed tomography (SPECT) has an important and well-validated role in the diagnosis of coronary artery disease\cite{gimelli2009stress}. The modality is widely used with close to nine million scans performed annually in the USA alone \cite{gowd2014stress}. Attenuation compensation (AC) is known to benefit visual interpretations of MPI by SPECT. However, AC typically requires an attenuation map, which is obtained must often on a dual-modality SPECT/CT system. \cite{bailey1987improved,patton2008spect}. This has multiple disadvantages, a key one being that multiple SPECT systems, including those in many community hospitals and most physician offices, mobile SPECT systems, and solid-state-detector-based SPECT systems, do not have a CT component \cite{technavio2021spectmarket}. Even when a SPECT system is available, acquiring CT images leads to increased costs, exposure to higher radiation dose, hardware complexity, greater training, regulatory concerns \cite{brenner2007computed,biermann2013there,sundaram2009role}, as well as the potential for diagnostic inaccuracies due to misalignment between SPECT and CT images \cite{saleki2019influence}. Thus, there is an important need for transmission-less AC methods that conduct AC with SPECT data only.

Given this need, multiple transmission-less AC methods have been proposed\cite{censor1979new,zaidi2003determination,krol2001algorithm,rahman2020fisher}, including those based on deep learning (DL) \cite{hagio2022virtual,chen2022direct,chen2022ct,chen2022cross,chen2023deep,yu2021physics,yu2023development,yu2024ctless}. A recently proposed scatter-window projection and DL-based transmission less AC method for SPECT (CTLESS) has shown particularly strong promise. The method demonstrated statistically non-inferior performance on the clinical task of cardiac perfusion defect detection, compared to a standard CT-based AC method (CTAC) with retrospective clinical data acquired on SPECT scanners from a single vendor \cite{yu2024ctless}. DL-based imaging methods for nuclear medicine have often been observed to perform sub-optimally on external datasets compared to the data from the same distribution used for training\cite{zech2018confounding,yasaka2018deep}. In a multicenter setting, SPECT scanners are often from different vendors. Therefore, for clinical translation, it is necessary to assess the generalizability of DL-based AC methods across SPECT scanners from different vendors \cite{jha2022nuclear}.
Also, it is important that such an assessment be performed on the clinical task, as previous studies have shown that evaluation of DL-based methods using task-agnostic figures of merit, such as root mean square error and structural similarity index measure, may yield discordant interpretations when compared to evaluations based on clinical tasks \cite{yu2020ai, yu2023need, li2021assessing, badal2019virtual}. The promising performance of CTLESS on the clinical task motivates further evaluation of the generalizability of CTLESS on the myocardial perfusion defect-detection task across SPECT scanners from different vendors.

Assessing the generalizability of a DL-based imaging method across SPECT scanners in a clinical study would typically require administering radiotracers multiple times to a patient, followed by imaging across different scanners. In practice, this process would result in substantial radiation exposure to patients, be expensive and time-consuming, and have multiple logistical challenges. Moreover, evaluating the CTLESS method on the myocardial perfusion defect-detection task requires knowledge of the ground truth for perfusion defects, which is generally unavailable in clinical images. While physical phantoms provide an alternative, they are restricted by the inability to represent the diverse anatomical and physiological characteristics of patient populations. In this context, \textit{in silico} imaging trials offer a compelling mechanism to perform early-stage evaluations and identify promising methods for clinical evaluation. In these trials, the patient population and imaging systems are simulated to model a clinically realistic imaging process. Simulating the patient population enables the modeling of variability in physiology and anatomy in patient populations while simultaneously knowing the ground truth \cite{badano2021silico,abadi2020virtual,jha2021objective}. Given these advantages, in this study, we conducted a virtual imaging trial, titled \textit{\underline{i}n \underline{s}ilico} \underline{i}maging \underline{t}rial to assess \underline{gen}eralizability (ISIT-GEN). ISIT-GEN assessed the generalizability of CTLESS across SPECT scanners from different vendors on the clinical task of myocardial perfusion defect detection. ISIT-GEN was designed to realistically capture patient and imaging system variabilities while accurately simulating clinical imaging systems and protocols. The performance of CTLESS was compared with the standard-of-care CTAC method and a reconstruction method without AC (NAC).

\section{Methods}
\subsection{Study design}
\begin{figure}[h!]
\centerline{\includegraphics[width=0.56\columnwidth]{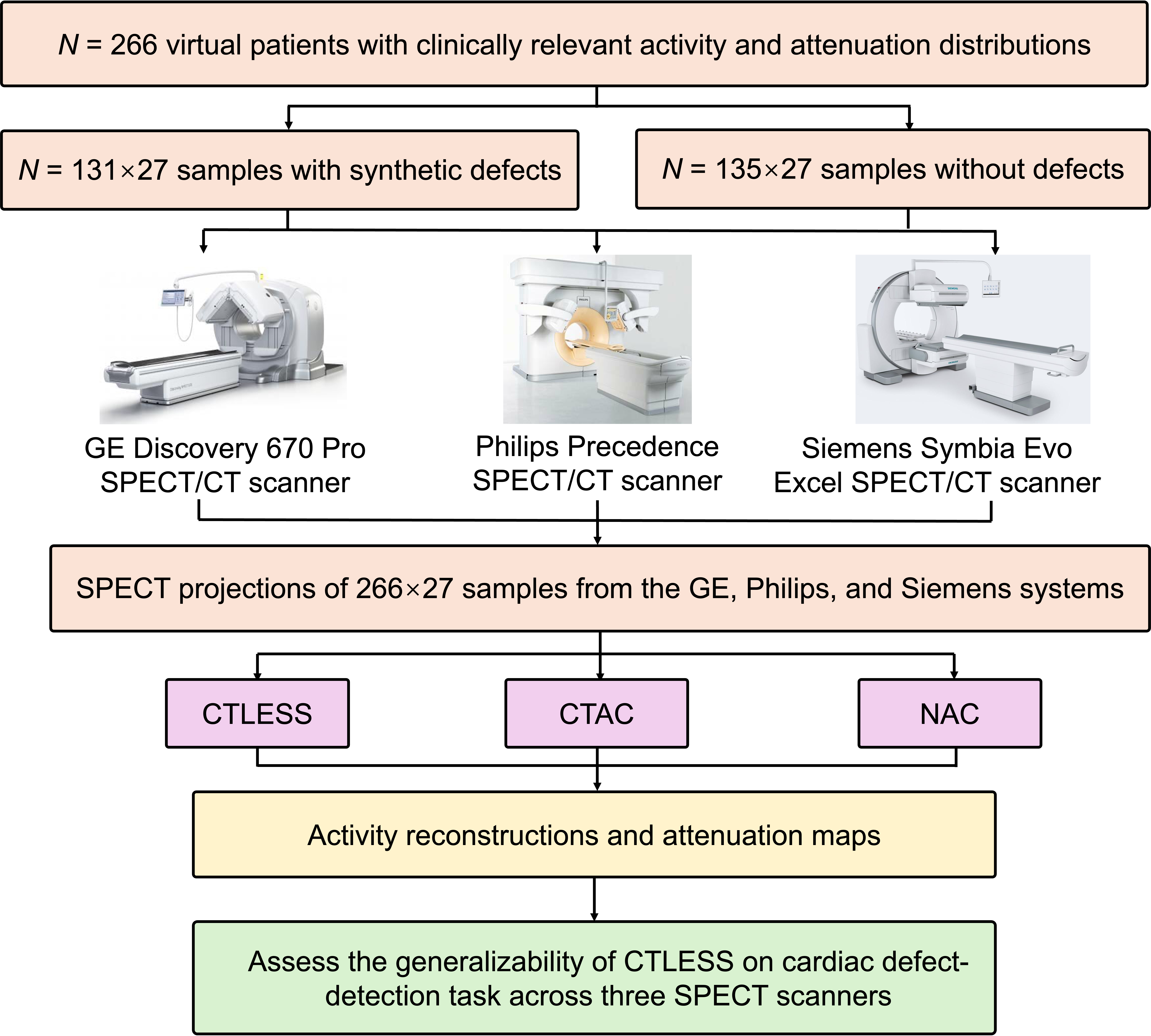}}
\caption{The overall design of the ISIT-GEN. The sources of images of the SPECT systems are the websites of GE HealthCare (https://www.gehealthcare.com), Philips Healthcare (https://www.medical.philips.com), and Siemens Healthineers (https://www.siemens-healthineers.com).}
\label{fig:chp6_workflow}
\end{figure}

ISIT-GEN aimed to assess the performance of the CTLESS method on the myocardial perfusion defect-detection task across SPECT scanners from different vendors. Fig.~\ref{fig:chp6_workflow} shows a visual representation of the overall trial design. This study was conducted at Washington University in St. Louis.
The primary objective of ISIT-GEN was to assess the inter-scanner generalizability of CTLESS on myocardial perfusion defect detection in an anthropomorphic model observer study.

\subsection{Trial population}
We collected anonymized data from $N = 266$ patients who underwent rest and stress MPI studies. CT images of those patients were acquired with the GE Optima CT 540 scanner component integrated on GE Discovery NM/CT 670 Pro systems. These CT images were used to generate $N = 266$ virtual patients. We segmented these CT images using TotalSegmentator \cite{wasserthal2023totalsegmentator}. To realistically simulate the physiological variations, we assigned clinically relevant activity uptakes in segmented regions across the patient population \cite{he2004mathematical}. The attenuation maps of these virtual patients were generated from the CT images, by converting the Hounsfield units to linear attenuation coefficients in cm\textsuperscript{-1} using a bi-linear model \cite{brown2008investigation}.

As mentioned earlier, the performance of CTLESS was evaluated on the myocardial perfusion defect-detection task. For this purpose, we designed 27 types of realistic cardiac perfusion defects with different extents, severities, and at different locations. Characteristics of these synthetic defects were similar to those in previous studies \cite{narayanan2001optimization,wollenweber1999comparison}. These defects were inserted into $N = 131$ virtual patients in the trial population, yielding $131 \times 27 = 3537$ samples, referred to as defect-present samples. For the remaining N = 135 virtual patients, we also generated $135 \times 27 = 3645$ samples without inserted defects (referred to as defect-absent samples), although samples generated from the same virtual patients were identical. Following this procedure, we simulated a virtual patient population with clinically relevant anatomical and physiological variability.

\subsection{Imaging protocol}
We simulated a clinical scenario in which patients received a one-day stress MPI protocol with a single intravenous injection of \textsuperscript{99m}Tc-sestamibi. SPECT projections were generated using SIMIND, a well-validated Monte Carlo simulation software for SPECT\cite{ljungberg1989monte,toossi2010simind}. We simulated three clinical SPECT scanners, including a GE Discovery 670 Pro SPECT/CT scanner, a Philips Precedence SPECT/CT scanner, and a Siemens Symbia Evo Excel SPECT/CT scanner. Projections were acquired at 30 angular positions with a right anterior oblique to left posterior oblique orbit. For all virtual patients, we collected projections in both photopeak (126-154 keV) and scatter (114-126 keV) windows from these three scanners. The scanning time was adjusted to ensure at least five million total counts were detected from the thoracic region within the photopeak window.


\subsection{Implementation of CTLESS and comparison AC methods}
The CTLESS method was trained with retrospective data obtained from $N = 508$ patients who underwent one-day rest and stress MPI studies on the GE Discovery 670 Pro SPECT/CT system in a previous study\cite{yu2024ctless}. The CTLESS method estimated attenuation maps of virtual patients and then reconstructed photopeak projections using an ordered-subsets expectation maximization (OSEM) technique with the estimated attenuation maps for AC \cite{merlin2018castor}. The CTAC method reconstructed photopeak projections using the same OSEM-based approach with AC using CT-derived attenuation maps. NAC-based images were obtained using the same OSEM-based approach without AC. Following the clinical protocol, reconstructed activity maps were reoriented into short-axis and filtered using a Butterworth filter.

\subsection{Task-specific information extraction}
We objectively evaluated the generalizability of CTLESS across SPECT scanners on the defect-detection task using an anthropomorphic model observer. In this study, the defect location was known, while defect extents and severity levels varied, and the background was also varying. In previous MPI SPECT studies, for this detection task with similar defects, it has been observed that the channelized Hotelling observer (CHO) with rotationally symmetric frequency (RSF) channels emulates human observer performance \cite{narayanan2001optimization,wollenweber1999comparison}. Thus, we used this anthropomorphic model observer with four RSF channels \cite{narayanan2001optimization,wollenweber1999comparison}.

To apply this CHO, we extracted a $32 \times 32$ region from the middle 2-dimensional slice of the short-axis images, ensuring that the defect centroid was positioned at the center of the extracted region. Using a leave-one-out strategy, the observer template was estimated and then test statistics were calculated for each defect-present and defect-absent sample. Based on these test statistics, we plotted receiver operating characteristic (ROC) curves using the LABROC4 program \cite{metz1998maximum,metz1986roc}.

\subsection{Figures of merit and statistical analysis}
The primary objective of ISIT-GEN is to assess the inter-scanner generalizability of CTLESS on the clinical task of myocardial perfusion defect detection. For this objective, we calculated the area under the ROC curve (AUC) yielded by CTAC, CTLESS, and NAC. Based on a power analysis, we derived the test data size to be $N = 3645~(135 \times 27)$ defect-absent and $N = 3537~(131 \times 27)$ defect-present samples. With these test samples, AUC values with 95\% CIs were calculated using a non-parametric strategy that accounted for the correlated nature of the data \cite{delong1988comparing,obuchowski1997nonparametric}.

For all statistical tests in ISIT-GEN, a \textit{p}-value $< 0.05$ was used to infer statistical significance. To account for multiple hypothesis testing, Bonferroni correction was applied.

\section{Results}

\begin{figure}[h!]
\centerline{\includegraphics[width=0.85\columnwidth]{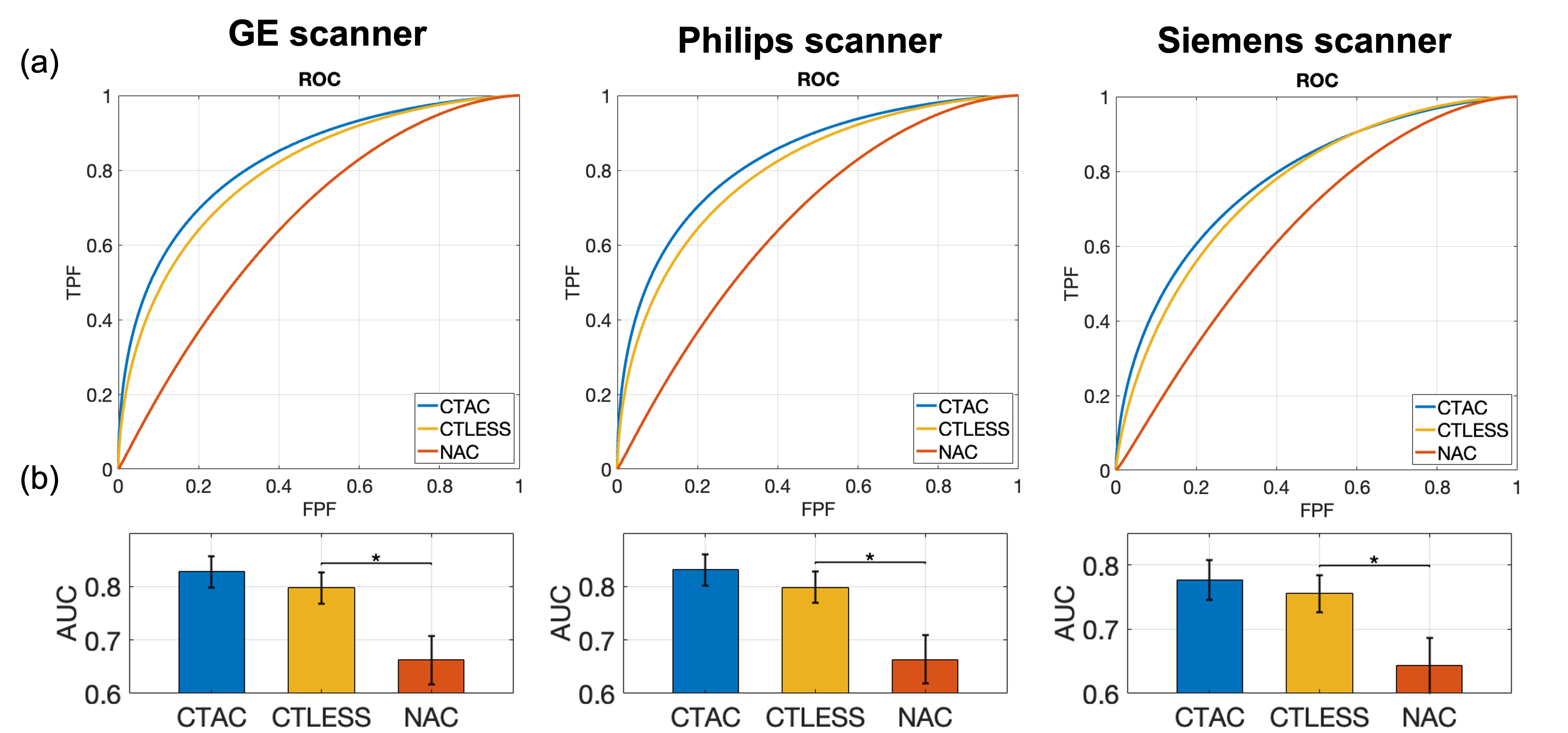}}
\caption{(a) ROC curves and (b) AUC values yielded by CTAC, CTLESS, and NAC across three considered scanners.}
\label{fig:chp6_AUC_ROC}
\end{figure}

Fig.~\ref{fig:chp6_AUC_ROC}a shows the ROC curves obtained by CTLESS, CTAC, and NAC methods on the defect-detection task. We observed that the ROC curves obtained by CTLESS were consistently close to those obtained by CTAC and outperformed the NAC method across three SPECT scanners. In addition, we observed that AUC values yielded by CTLESS were close to those yielded by CTAC and significantly higher than those yielded by NAC ($p<0.05$) across three SPECT scanners (Fig.~\ref{fig:chp6_AUC_ROC}b). 

\section{Discussion and conclusion}
The reliability and robustness of DL-based medical imaging methods can be compromised due to previous unseen and out-of-distribution samples, which may arise from differences in patient populations, scanners, imaging protocols, and reconstruction parameters. Evaluation of the generalizability of DL-based medical imaging methods is critical to assess their potential for clinical translation. In ISIT-GEN, we specifically assessed the inter-scanner generalizability of a scatter-window and DL-based AC method for MPI-SPECT, namely CTLESS, across SPECT scanners from three different vendors, with the performance evaluated on the clinical task of myocardial perfusion defect detection.

Fig.~\ref{fig:chp6_AUC_ROC}a demonstrates that for all three SPECT scanners, the ROC curves obtained by the CTLESS method were consistently close to those obtained by the CTAC method and outperformed the NAC method. As shown in Fig.~\ref{fig:chp6_AUC_ROC}b, the AUC values yielded by CTLESS were consistently close to those yielded by CTAC and significantly higher than those yielded by NAC.
These observations indicated that the CTLESS method generalized well across the three SPECT scanners from different vendors.

We evaluated the inter-scanner generalizability of CTLESS in an \textit{in silico} imaging trial, which not only provided known ground truth for evaluations on the myocardial perfusion defect-detection task but also simulated clinically relevant patient population and clinical SPECT systems \cite{badano2021silico,abadi2020virtual,jha2021objective}. More importantly, \textit{in silico} imaging trials provide a mechanism where the same virtual patient can be scanned multiple times on different scanners, overcoming the practical challenges in clinical trials. To ensure a high degree of clinical realism, ISIT-GEN was meticulously designed throughout the trial pipeline. The anatomical characteristics of the virtual patients were derived from CT scans of real patients, while the physiological characteristics were based on clinical data \cite{he2004mathematical}. Clinical SPECT systems were simulated using well-validated Monte Carlo simulations, with clinically relevant parameters and an imaging protocol aligned with the clinical guidelines \cite{dorbala2018single}. Additionally, we evaluated the performance of CTLESS on the clinical task of myocardial perfusion defect detection, thus making the study results clinically relevant and strengthening the clinical significance of the study.

While ISIT-GEN was carefully designed to ensure a high degree of clinical realism, this study has several limitations. First, the simulations may not have been able to capture all aspects of variations in patient physiology and anatomy. Second, we assumed a static distribution of the tracer across acquisitions and did not simulate the pharmacokinetics of the tracer. Future studies could incorporate patient-specific \textsuperscript{99m}Tc-sestamibi pharmacokinetic data to enhance the realism of virtual patient simulations. Additionally, while the synthetic defects used in this study were designed to be clinically relevant, an ideal evaluation would involve real perfusion defects at unknown locations in the heart. However, the results from this study motivate clinical evaluation.

In conclusion, we conducted an \textit{in silico} imaging trial, titled ISIT-GEN, to assess the inter-scanner generalizability of CTLESS for MPI-SPECT on the myocardial perfusion defect-detection task. Results from ISIT-GEN demonstrate that the CTLESS method generalized well across a GE Discovery 670 Pro SPECT/CT scanner, a Philips Precedence SPECT/CT scanner, and a Siemens Symbia Evo Excel SPECT/CT scanner, showing significantly better performance than the NAC method on the defect-detection task. These results motivate further clinical evaluation of the CTLESS method.

\acknowledgments 
This work was supported in part by National Institute of Biomedical Imaging and Bioengineering of National Institute of Health (NIH) under grant number R01-EB031051, and R01-EB031962, and the Bradley-Alavi Student Fellowship from the Education and Research Foundation for Nuclear Medicine and Molecular Imaging (ERF) of the SNMMI.

\bibliography{ref} 
\bibliographystyle{spiebib} 

\end{document}